\documentclass[reprint,https://v2.overleaf.com/project/5ba006002181f90dc7afddb8]{revtex4-2}
\usepackage{CJK}
\usepackage{amsmath}
\usepackage{amssymb}
\usepackage{graphicx} 
\usepackage{dcolumn}  
\usepackage{bm}       
\usepackage[caption=false]{subfig}
\usepackage{hyperref}
\usepackage{float}
\usepackage{xcolor}
\hypersetup{colorlinks=true, pdfstartview=FitV, linkcolor=blue, citecolor=black, plainpages=false, pdfpagelabels=true, urlcolor=blue}
\usepackage[all]{hypcap}
\usepackage[normalem]{ulem}
\usepackage{tabu}

\begin{document}

\title{Controlling dynamics of stochastic systems with deep reinforcement learning}

\author{Ruslan I. Mukhamadiarov}
\affiliation{Arnold Sommerfeld Center for Theoretical Physics and Center for NanoScience, Department of Physics, Ludwig-Maximilians-Universität München, Theresienstrasse 37, D-80333 Munich, Germany}


\begin{abstract}
A properly designed controller can help improve the quality of experimental measurements or force a dynamical system to follow a completely new time-evolution path. Recent developments in deep reinforcement learning have made steep advances toward designing effective control schemes for fairly complex systems. However, a general simulation scheme that employs deep reinforcement learning for exerting control in stochastic systems is yet to be established. In this paper, we attempt to further bridge a gap between control theory and deep reinforcement learning by proposing a simulation algorithm that allows achieving control of the dynamics of stochastic systems through the use of trained artificial neural networks. Specifically, we use agent-based simulations where the neural network plays the role of the controller that drives local state-to-state transitions. We demonstrate the workflow and the effectiveness of the proposed control methods by considering the following two stochastic processes: particle coalescence on a lattice and a totally asymmetric exclusion process. 
\end{abstract} 

\maketitle

\section{Introduction}
Gaps in theoretical knowledge can seriously compromise the control efforts since control application relies on dynamical equations that describe the time evolution of the system \cite{trentelman2002control, glad2018control}. At the same time, a variety of dynamic aspects of a system can be learned when trying to exert control over it. A representative example of how theory and control can benefit from each other is aerospace exploration, where the successful design of propulsion, guidance, and landing control systems was accomplished after failure-driven improvements in understanding of aerodynamics, combustion, high-pressure physics, and other fields of science. Furthermore, application of control techniques can drastically increase the performance of experimental setups that use feedback loops \cite{Bechhoefer_2021}, e.g., atomic force microscope (AFM) \cite{PhysRevLett.56.930, 10.1063/1.1143970, RevModPhys.75.949}, and with this open new doors for cutting-edge scientific exploration. The central goal of control theory is to design a control scheme or a controller that is capable of altering the dynamics of the system or driving state-to-state transitions in an effective manner \cite{RevModPhys.77.783}. These tasks are achieved by choosing control setups that minimize the error between the controller's output and the desired signal. A typical controller consists of feed-forward, feedback elements, PID (proportional-integral-derivative) controllers, or cascades of these elements. If equations that describe the dynamics of the system are known, one of the most efficient ways to instill a control law on a system is to use established methods from the optimal control, which tune the controller parameters and then evaluate the controller's performance through a certain index, e.g., action, thermodynamic potential, survivability \cite{kirk2004optimal, sethi2019optimal}.

In practice, the ability of a controller to fulfill its purpose may be seriously impacted by incomplete knowledge of the system's dynamics and by uncertainties that are not directly accounted for by the dynamical model of the system. The latter often comes from environmental disturbances, aging of system components, or statistical noise (e.g., from sensor readings) and can be tackled by minimizing the mean squared error between predicted and observed internal states with the help of the Kalman filter and its variations \cite{10.1115/1.3662552, welch1995introduction}. As for explicitly missing information, robust control methods have been developed to account for the gaps in the knowledge by minimizing the error between partially known and the ``true" dynamics of the system \cite{zhou1998essentials,green2012linear}. Other fairly recent approaches that attempt to learn about the system's dynamics and reduce the uncertainties while simultaneously exerting control over it are adaptive control and reinforcement learning \cite{aastrom1995adaptive, tao2003adaptive, kaelbling1996reinforcement, meyn2022control}.

A serious limitation of the methods mentioned above is a prohibiting number of equations and their control parameters, which need to be optimized to achieve control over complex systems. However, with the rising popularity of machine learning and the realization that feed-forward neural networks can serve as universal approximators \cite{scarselli1998universal}, a new powerful approach that is called deep reinforcement learning gained considerable attention in the scientific community \cite{mnih2013playing, arulkumaran2017deep}. The approach combines the use of neural networks and concepts that have already been developed from reinforcement learning that enable efficient learning of the system dynamics through interactions of the agent with the environment. To mention some of its applications, in statistical mechanics of non-equilibrium systems, deep reinforcement learning is used for exerting control on epidemic systems \cite{10.1007/978-3-030-67670-4_10, jain2024sirrlreinforcementlearningoptimized}, for studying the problem of urban traffic \cite{lin2018efficientdeepreinforcementlearning}, and for studying cooperation dynamics of predator-prey models \cite{WANG2020100815, doi:10.1098/rsif.2020.0639}. In biophysics, reinforcement learning has been used to devise efficient navigation strategies \cite{PhysRevE.107.055105, doi:10.1098/rspa.2022.0118, rando2025qlearningtemporalmemorynavigate, Caraglio_2024} and state-to-state transition protocols for active matter \cite{Casert2024}, for finding optimal self-assembly protocols \cite{PhysRevE.101.052604}, and for studying communicating matter \cite{Grauer_2024, tovey2024emergencechemotacticstrategiesmultiagent} and smart microswimmers \cite{tovey2024swarmrlbuildingfuturesmart, Cichos2020}. In quantum systems, reinforcement learning has been used to enhance quantum control techniques \cite{PhysRevX.8.031086}, to improve the performance of quantum simulations \cite{Niu2019}, and to optimize quantum variational circuits \cite{Lockwood_Si_2020}. Outside of physics, reinforcement learning is best known for finding efficient game strategies that outperform world champions \cite{silver2018general, berner2019dota}, for training robots \cite{gu2017deep, ibarz2021train}, language recognition models \cite{serban2017deep, uc2023survey}, computer vision models \cite{voulodimos2018deep, le2022deep}, etc.

Deep reinforcement learning has brought us closer to a control engineer's dream, where one states a desire, turns on the controller, and lets the controller figure out what to do \cite{Bechhoefer_2021}. However, it is still not clear whether the deep reinforcement learning method can be successfully applied to control multi-agent stochastic systems. Would a targeted control be at all effective on a microscopic scale in the presence of random forces, and to what degree must the underlying dynamics of stochastic systems be changed for a controller to achieve specific objectives?

In this work, we investigate these questions by proposing a new simulation method that controls the dynamics of stochastic systems with the help of deep reinforcement learning. The idea behind the method is to use artificial neural networks as a controller that computes transition probabilities between different local states for individually selected agents. In order to test the effectiveness of our control method, we simulate the stochastic dynamics of two different processes on a regular lattice -- the particle coalescence and the driven particle transport with exclusion. We find that replacing the original reaction scheme with a neural network controller that reshapes transition probabilities is reminiscent of developing new interparticle interactions that help to achieve the desired dynamical behavior. At the same time, by the design of our simulation method, sequentially selecting random agents for the update in multi-particle systems undermines the effectiveness of the control measures since the agents cannot react to the ever-changing environment until they are picked again for the update. Finally, we were able to achieve more complex control objectives by employing a compound reward structure, as we show in our studies of the heterogeneous traffic problem.

\section{Model}

Before we demonstrate how an artificial neural network can act as a controller, we first draw connections between reinforcement learning and control theory and speculate that solving the reinforcement learning problem can be equivalent to finding a control law for the specific control task at hand. The reinforcement learning goal concludes in finding a policy $\pi$ that maximizes a cumulative reward $R$:
\begin{equation}
    R = \sum_{t=0}^T \gamma^t r_t,
\end{equation}
where $T$ is the duration of a single training episode, $r_t$ is an instantaneous reward, and $0 \le \gamma < 1$ is a discount factor that ensures that the sum does not diverge. In analogy with a control law, reinforcement learning policy $\pi$ acts as a map that takes as an input a state of the system $s$ and maps it to a probability distribution over different possible actions that the agent can take $\pi: s\to \{P (a|s)\}$.

The state $s$, the action $a$, the corresponding immediate reward $r$, and the next state $s'$ are the key components that together constitute the Markov Decision Process (MDP) framework that lies at the core of the reinforcement learning. The action that brings the system from state $s$ to another state $s'$ can be compared with a control input vector $u(t)$ that manipulates the dynamics of some system $\dot{x} = F(x,t) + u(t)$, and the instantaneous reward $r$ that is attributed to the state-action pair $(s,a)$ can be related to the cost-to-go in the adaptive control, with total reward $R$ resembling the optimization cost function. Similarly to looking for a control law that optimizes the total cost, the task of reinforcement learning is to find the best policy $\pi$ that, within the MDP framework, allows the attainment of a specific goal encoded into an appropriately chosen set of instantaneous rewards. If the dynamics of the system is known, i.e., one knows the transition probability $p(s',r|s,a)$, where $s'$ is the next state, then finding the optimal policy comes down to solving Bellman optimality equations \cite{Sutton1998}. However, finding a control law for a system with known dynamics can also be done with other well-developed methods from control theory, such as optimal control.

One of the key strengths of the reinforcement learning method is its ability to find optimal solutions to the control problem when we know nothing about the rules that prescribe the system's evolution. For these scenarios, one can collect the state transition experiences, record them into $\{s,a,r,s'\}$ table, and then use the bootstrapping methods to approximate the optimal policy \cite{osband2016deep, kumar2019stabilizing}. The downside of such an approach is the inoperable size of the experience tables and huge sets of coupled Bellman equations that result from these tables, which, until recently, severely encumbered practical applications of reinforcement learning to large-scale systems \cite{kaelbling1996reinforcement}. Only after substantial advances in machine learning did it become possible to apply the ideas from reinforcement learning for large systems by replacing the need to solve innumerable sets of equations with a neural network parameter optimization procedure.

\begin{figure}[t!]
    \centering
    \includegraphics[width=\columnwidth]{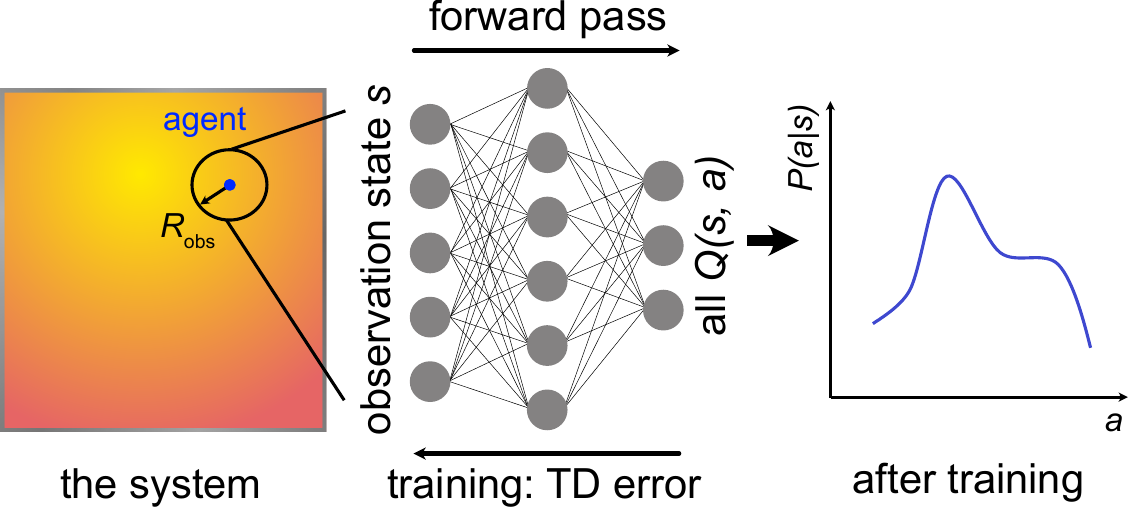}
    \caption{Schematic illustration of how a neural network drives local state-to-state transitions in the agent-based simulations. The observation state $s$ of the agent is provided to the artificial neural network as an input vector, which in turn produces $Q$-values for each action $a$. As such, the size of the input layer of our neural network is set by the size of the observation state vector $s$, and the output layer size corresponds to the size of the action space $a$. The temporal-difference (TD) error is computed with Eq.~\eqref{eqn:Bellman} and backpropagated to update the network parameters. After the neural network training is done, the network parameters become fixed, and the $Q$-values are mapped to transition rates $P(a|s)$ using Softmax function in Eq.~\ref{eqn:Softmax}}\label{fig:schematics}
\end{figure}

To investigate how effectively stochastic systems can be controlled with deep reinforcement learning, we introduce the agent-based simulation method, which is a combination of a familiar random sequential update scheme and a trained neural network that drives local state transitions. In contrast to typical stochastic simulations where the system's dynamics is specified by hard-coded transition rates, the transition rates in our method are determined by an artificial neural network. As it is shown in Fig.~\ref{fig:schematics}, we provide an observation state as an input to the feed-forward neural network and interpret the output as $Q(s,a)$ values that are then converted to transitional probabilities $P(a|s)$. However, before the neural network can do such mapping and fulfill desired control objectives, it needs to be trained first.

As such, we separate the training and the post-training phases, where in the training phase we keep adjusting the neural network parameters by using the established $Q$-learning algorithm from reinforcement learning \cite{watkins1992q, hester2018deep}, and only then we perform all necessary measurements by running simulations with fixed neural network parameters. The essence of the $Q$-learning algorithm consists of approximating the Bellman equations for the so-called $Q$-values -- the values that serve as a measure of how good it is to take action $a$ while being in the state $s$. The $Q(s,a)$ values and the network parameters are then updated by backpropagating the temporal-difference (TD) error, which is the right-hand side of the following equation \cite{Sutton1998}:
\begin{equation}\label{eqn:Bellman}
    \begin{aligned}
       Q_{t+1}(s,a) &- Q_t(s,a) =\,  \\
       &\alpha \left(r + \gamma\, \underset{a}{\max}\, Q_t(s',a) - Q_t(s,a) \right) 
    \end{aligned}
\end{equation}
where $\alpha$ is a learning rate, $r$ is an immediate reward that agent receives for doing action $a$ in state $s$, and $s'$ is the next state. The computed error in the brackets estimates a future reward gain for performed action $a$ from state $s$. A positive value of the error increases the $Q$-value of a state-action pair $(s,a)$, thus reinforcing the network to repeat this pattern and choose the same action $a$ if it receives the state $s$ again.

At the beginning of the training, the neural network itself possesses no information about the control objectives and which state transitions are ``good" or ``bad". Therefore, the training starts with the flat distribution of $Q(s,a)$ values, and the instantaneous rewards $r$ play a big role in reshaping this distribution during the training. To update the neural network parameters that map system's state $s$ to $Q(s,a)$-values for possible actions $a$, we collect transition experiences into a $\{s,a,r,s'\}$ table and, after every $N$ transitions, we update the neural network parameters by backpropagating the temporal-difference error computed for collected experiences using Eq.~\ref{eqn:Bellman}. To efficiently explore the state-action space, we let the network choose random actions at the early stages of the training, while later on, we shift towards exploiting the accumulated knowledge and encouraging the network to take actions that are expected to bring more reward. Such training scheme is called exploration-exploitation training $\varepsilon$-$greedy$ algorithm, and it can be expressed as:
\begin{equation}\label{eqn:vareps}
    a = 
    \begin{cases}
        \,\underset{a}{\arg \max} Q(s,a) \hspace{0.5cm} \text{with probability } 1-\varepsilon \\[8pt]
        \,\text{random action} \hspace{0.8cm} \text{with probability } \varepsilon
    \end{cases}
\end{equation}
where $\varepsilon = e^{-t/\tau}$ decays exponentially with time, and the decay rate $\tau$ is a parameter that regulates how quickly the network transitions from choosing random actions for the sake of exploration to choosing the best action at a given state $s$.

Once the cumulative temporal-difference error, or simply the loss, reaches some saturation value in the course of the training, we consider the training phase to be over. As our next step, we fix the parameters of the neural network and move to the post-training phase, where the actions are now sampled using transitional probabilities $P(a|s)$ that are obtained from $Q$-values using Softmax function: 
\begin{equation}\label{eqn:Softmax}
    P(a|s) = \frac{\exp (-\,Q(s,a))}{\underset{a}{\sum} \exp (-\,Q(s,a))}
\end{equation}
where the sum is performed over all possible action $a$, i.e., over all nodes in the output layer of the neural network that performs the state-action mapping (Fig.~\ref{fig:schematics}). For situations when produced $Q$-values lie all too close to each other, one could use $\Delta Q(s,a) = Q(s,a) - \langle Q(s)\rangle_a$ in Eq.~\ref{eqn:Softmax} to highlight the differences of taking different actions. If the system contains multiple agents, we continue picking them randomly, and, for each picked agent, we provide their state to a neural network, sample the action, and update the agent's state. In our simulations, all agents share the same artificial neural network. For a detailed implementation, see the code in our GitHub repositories \cite{my_repo, jonas_maert_github}.

We want to highlight the importance of immediate rewards $r$ that enter Eq.~\ref{eqn:Bellman} that updates $Q$-values. Conceptually, the assignment of instantaneous rewards $r$ for the state-action pairs $(s,a)$ sets the initial direction for the gradient descent optimization procedure, which explores the landscape of the high-dimensional $Q(s,a)$ function. Since the latter is quite complex, properly choosing the immediate rewards that allow achieving a desired state-to-action mapping is one of the hardest challenges in reinforcement learning. Quite often, a seemingly reasonable choice of rewards can lead to unexpected, undesired system behavior. It is therefore advised to perform as many sanity checks and assisting measurements as possible to confirm that the immediate rewards indeed provide a correct incentive towards desired system dynamics. In this paper, we demonstrate how the different choices of instantaneous rewards affect the ability of a neural network controller to achieve desired control objectives.

\section{Simulations results}
\subsection{Simple Rewards: Particle Coalescence} 

We choose the particle coalescence process $A+A\to A$ on a lattice as the starting point of our exploration of how well deep reinforcement learning can control the dynamics of the stochastic processes with the simulation scheme we proposed. Among many implementations of particle coalescence, we choose the model where particles can only jump on next-neighbor lattice sites, and the lattice sites are allowed to host at most one particle, i.e., the lattice site occupation numbers are restricted to 0 and 1. If a particle jumps on an occupied site, the two particles merge, and only one remains. Starting from a fully occupied lattice, the coalescence process continues until only one particle is left in the system. If the particle hopping rates are fixed and symmetric, the number of particles in the system would be dropping with time asymptotically as $t^{-1/2}$ in a one-dimensional system or as $t^{-1} \log t$ in two dimensions \cite{Sawyer, bramson1980asymptotics}. We later refer to this asymptotic behavior as normal decay.

\begin{figure}[t!]
    \centering
    \includegraphics[width=\columnwidth]{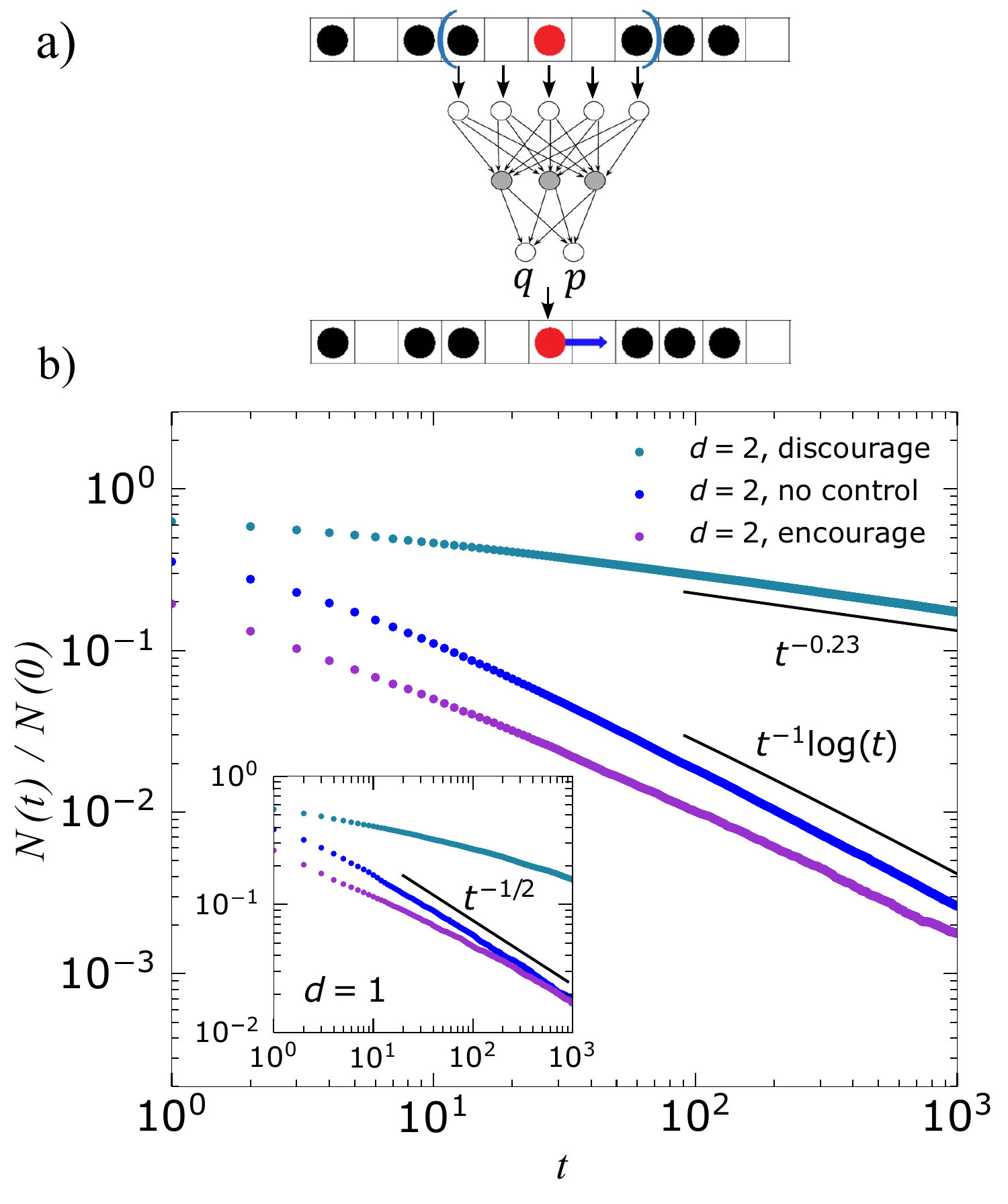}
    \caption{(a) Schematic illustration of how a neural network controls dynamics of the coalescence process in one dimension. The selected agent is marked in red, and the blue brackets show the size of the agent's observation state that is being provided to the network. The network computes $Q$-values, which, after training, can be mapped to hopping probabilities $q$ and $p$ with the Softmax function. (b) Time evolution of the particle number fraction for different control scenarios on $d=2$ and $d=1$ (inset) lattices. The simple reward choices of $-1$ or $+1$ have been used to train the networks to discourage or encourage coalescence events, respectively. The data was obtained by running simulations after the training was finished. The simulations were run on $L=1000$ and $100\times 100$ lattices with periodic boundary conditions. Each data point was obtained by averaging over $10$ independent simulation runs.}\label{fig:coalescence}
\end{figure}

The simplest controlling objective for a coalescence process that one can come up with is to force the coalescence dynamics to deviate from its normal asymptotic behavior by either speeding up the merging of particles or slowing it down. To achieve this, we choose immediate rewards to be $r=+1$ or $r=-1$ for every action that leads to a coalescence event if we want to encourage or suppress coalescence, respectively. As we show in Fig.~\ref{fig:coalescence} (a), for a selected agent, we pass to the neural network the agent's observation state $s$, which in one dimension is a $2R_{obs} + 1$ vector of lattice occupations numbers, with $R_{obs} = 5$ being the agent's observation radius. For simulations on a two-dimensional lattice, we choose the occupation numbers of the agent's four nearest neighbors to constitute the observation state $s$. The two possible actions for one-dimensional coalescence are jumping to the left or to the right nearest-neighbor lattice sites, while in two dimensions, the agent can jump in any of the four directions. Therefore, the neural network output vector size is limited to 2 for $d=1$ and to 4 for $d=2$ simulations, respectively. After receiving the observation state $s$, the neural network computes a corresponding $Q(s,a)$-value for each possible action $a$. Then, the agent either takes action with the biggest $Q$-value or takes a random action, following the $\ varepsilon$-$greedy$ algorithm in Eq.~\ref{eqn:vareps}. After the agent's state is updated, we record the old state $s$, jump direction $a$, reward $r$, and the new state $s'$ in the experience table. We use this information later to update the network parameters by computing the temporal-difference error in Eq.~\ref{eqn:Bellman} and backpropagating it across the network. The particles for the update are chosen randomly, and a single time step is considered to be over when each particle is picked at least once on average. A single training episode is considered to be over once there are less than $5\%$ lattice sites occupied or if the particle number stays unchanged over a very long time.

Before carrying out the measurements using a trained neural network, we perform a benchmark test, where we first run simulations with an untrained network that drives state-to-state transitions. As one can expect, an untrained network with randomly chosen parameters will be performing updates at random, disregarding any control objectives. This, in turn, leads to a normal particle decay, as shown in Fig.~\ref{fig:coalescence} (b). After the network has been trained to encourage particle merging, we observe a rapid drop in the particle number fraction right at the beginning of the simulation, which is followed by a slower decay, as particles need to traverse empty space to find the other particles, which finally succeeded by a crossover to a normal decay. We interpret the observed behavior as due to picked agents quickly merging with all the particles in its observation range, which leaves them surrounded with empty lattice sites and forces them to wander randomly until some other particle enters their observation range. In essence, an effort to control coalescence dynamics by accelerating the merging of particles boils down to the rapid merge of all particles within the observation patch that is followed by diffusion-driven merging of what one can imagine, ``quasi-particles'' with radius $R_{obs}$. Once the observation radius $R_{obs}$ becomes comparable with the size of the lattice, the neural network manages to collapse the whole system within a single time step \cite{jonas_braun}.

In contrast, teaching the neural network to try avoiding coalescence events leads to a significantly slower decay of particle number both on $d=1$ and $d=2$ lattices. The reason why we could not completely interrupt the coalescence process lies in the design of our simulation method, where agents are chosen randomly for an update. Due to random sampling, some particles can end up getting surrounded by other particles before they are picked again for an update, which is imminently followed by a coalescence event since particles always have to jump. Increasing the observation radius does not seem to resolve this issue completely, as the $Q$-values get less and less sensitive to the differences in lattice occupation numbers as the distance to a chosen particle increases. As a result, for our simple choice of immediate rewards, mapping Q-values to a set of transitional probabilities for the post-training simulations results in particles acquiring long-range repulsive interactions that quickly drop with distance. To develop strong long-range sensing, one possible solution would be to make an instantaneous reward explicitly dependent on the distance to the other particle. We investigate more complex choices of immediate rewards in the next scenario that we have considered.

\begin{figure*}[t!]
    \centering
    \includegraphics[width=2\columnwidth]{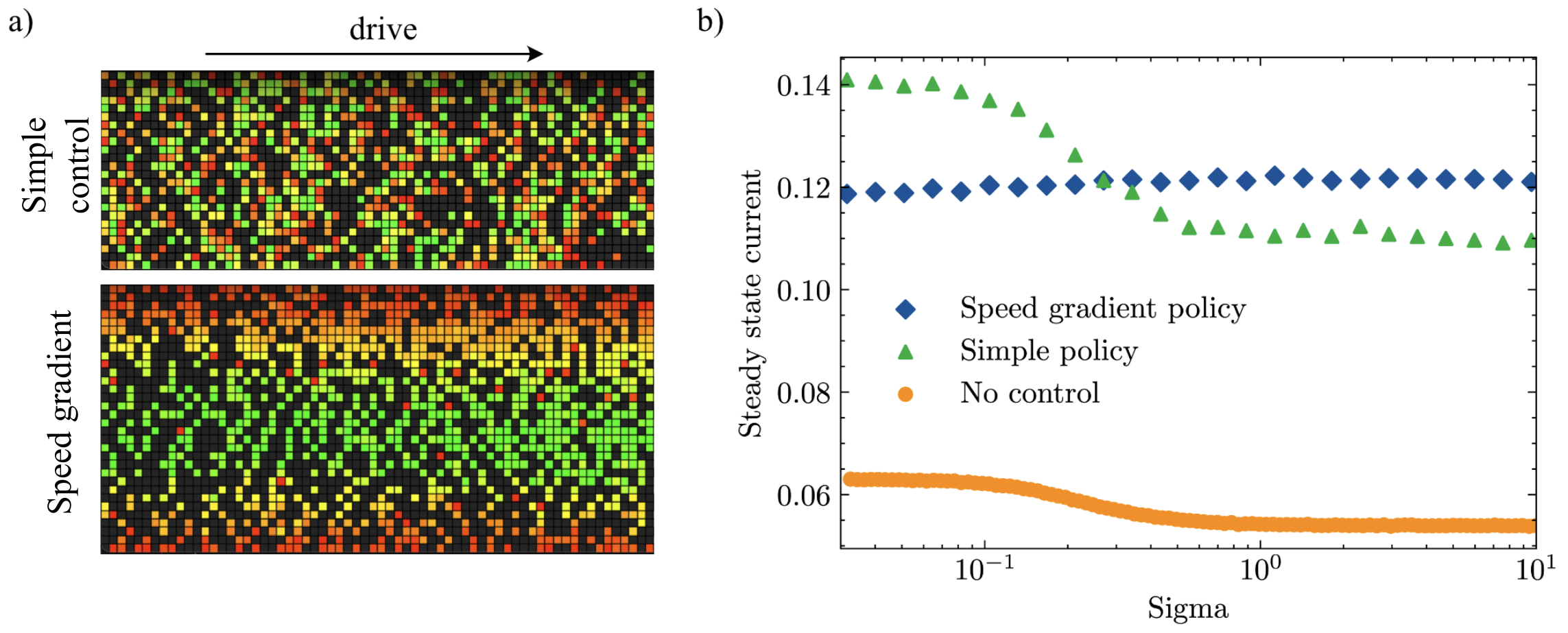}
    \caption{(a) Simulation snapshots of heterogeneous totally asymmetric exclusion process (TASEP), where green (light grey) particles have a higher overall jumping rate and red (dark grey) lower. The jump directions of each particle are selected by a neural network that makes a decision based on the particle's observation state. Here are two scenarios considered: a simple control scenario, where only successful forward jumps are rewarded, and a speed gradient scenario, where in addition to rewarding forward jumps, particles are also encouraged to separate into lanes based on their ``speed". The lattice sizes are $128\times 24$ and $128\times 32$, respectively. (b) The global current in heterogeneous TASEP obtained for different control policies in steady state and plotted against the standard deviation $\sigma$ in particle overall jumping rate $\nu \sim \mathcal{N}(0.5,\sigma^2)$ that serves as a measure of heterogeneity. The simulations were performed on half-filled lattices with periodic boundary conditions. Each data point has been obtained after averaging over 800 independent realizations. Adapted from Jonas Märtens Bachelor Thesis~\cite{jonas_maertens}.}\label{fig:tasep}
\end{figure*}

\subsection{Complex Rewards: Heterogeneous TASEP} 
The next stochastic process, the dynamics of which we seek to control, is called the totally asymmetric exclusion process (TASEP), and it is arguably one of the most paradigmatic models for studying the non-equilibrium transport \cite{zia2011modeling}. To put it briefly, a totally asymmetric exclusion process is a stochastic process, where a collection of hard-core interacting particles can hop on nearest-neighbor unoccupied lattice sites. The asymmetry in jumping rates comes from the presence of the drive that makes the probabilities to jump along and against the drive unequal, with this generating a global particle current in the direction of the drive. In the limiting case of infinite drive strength, all jumps against the drive become prohibited, and the process is called totally asymmetric. For a long time, it has been known that particle transport can drop significantly in the presence of inhomogeneities, such as slow bonds or slow particles \cite{Janowsky1994, borodin2009two}. In the one-dimensional case, the presence of such defects can trigger phase separation into a congested, high-density phase and a low-density phase, with a real-life example of traffic congestion due to some road bottlenecks that can also stem from accidents. In higher dimensions, the traffic problem that stems from the heterogeneity of a transport channel can be partially averted by a diffusive motion of agents in the direction(s) transverse to the drive.

Seeking to address a traffic problem in a heterogeneous environment of fast and slowly moving agents, we are interested in employing deep reinforcement learning to search for control strategies that maximize the global current in such a heterogeneous system. To this end, we consider a modified version of a totally asymmetric exclusion process, where each particle, on top of $1/2, 1/4, 1/4$ jumping rates in forward, up, and down directions, has another rate $\nu \sim \mathcal{N}(1/2,\sigma^2)$ that prescribes whether the particle would move at all or not. This additional rate that is being awarded to each moving particle independently from truncated normal distribution ($\nu \in [0;1]$) emulates the situation in road traffic, where different cars have different driving capabilities. To exert control on the dynamics of such a system, we pass a $5\times 5$ observation patch around some randomly picked ``car'' to a neural network and ask the network to choose the direction for the next jump.

In our attempt to maximize the current of particles along the drive in heterogeneous TASEP, we have explored two control strategies: a simple policy and a speed gradient policy (see Fig.~\ref{fig:tasep}(a)). In the case of a simple policy, for any successful jump along the drive, we give the network a reward $r=+1$. If the network chooses to jump on an already occupied lattice site, such an update goes to waste due to the exclusion, and we punish the selected action with $r=-1$ reward. For a speed gradient policy, inspired by fast-slow lane separation on the highways, in addition to a simple set of rewards used for simple policy, we encourage faster particles to move towards the channel's center and slower particles towards the edges of the channel that are connected with a periodic boundary. Specifically, we perform a linear mapping between the particle's ``speed" $\nu$ and its perpendicular position in the transport channel, increasing the reward linearly as the particle moves towards its designated lane. The other forms of reward function that we have tried for the speed gradient policy, as well as other control policies that encourage local clustering of particles with similar speeds, can be found in Jonas Märtens Bachelor Thesis~\cite{jonas_maertens}.

After training the neural networks, we perform particle current measurement for different degrees of the system's heterogeneity, which we represent with the standard deviation $\sigma$ that enters particle speed distribution $\nu \sim \mathcal{N}(1/2,\sigma^2)$. To estimate the degree of improvement, we first run simulations for an uncontrolled scenario, where the untrained network chooses the jump directions randomly. As we show in Fig.~\ref{fig:tasep} (b), in the absence of control, the current lies pretty low, converging to theoretically predicted 2$d$-TASEP value for homogeneous jumping rates ($\sigma \ll 1$). The lower current values for a high degree of heterogeneity can be explained by the spontaneous formation of short-lived blockages of slow particles that impede the overall motion. A trained neural network that has learned to follow a simple policy does not seem to dissolve these blockages, but the network forces particles to always jump forward if the next lattice site is empty and enables particles to go around the obstacles, which is reflected in a significantly higher parallel particle current. Finally, enforcing the speed gradient policy seems to lead to better performance for highly heterogeneous systems with high particle jumping rates variance ($\sigma > 1$). However, for homogeneous TASEP ($\sigma \ll 1$), the simple policy seems to produce a higher parallel current and outperform the speed gradient policy. Therefore, while for heterogeneous environments, the particles seem to benefit from learning to phase-separate into faster and slower lanes, for relatively homogeneous systems, according to the results of our simulations, a simple policy where particles learn to always jump forward when possible seems to be the most effective strategy to maximize the particle transport. While these are relatively obvious conclusions that do not require running stochastic simulations, the fact that our efforts to control the traffic dynamics with deep reinforcement learning nevertheless led us to the same conclusions brings more trust to the proposed method.

\section{Conclusion}

In this work, we have introduced a method for controlling the dynamics of stochastic systems using deep reinforcement learning. We showed how the reinforcement learning method is related to control theory and how the use of neural networks together with $Q$-algorithm automates the task of tuning a controller to achieve a certain control objective. Through our investigation, we found that deep reinforcement learning is capable of modifying the dynamics of stochastic systems by direct intervention and modification of each individual agent's local reaction scheme. We also found that the outcome of the training strongly depends on the choice of the initial rewards, which sometimes can have a non-intuitive functional form. This is related to a common pitfall of using deep reinforcement learning, where a suboptimal choice of initial gradient descent vector can lead to an optimizer being stuck in some local minima. Finally, the simplicity of the stochastic processes that we have considered has enabled us to take the first steps toward understanding the consequences of applying control to stochastic systems using artificial neural networks. However, this simplicity was also a major limiting factor in exploring the potentially far-reaching capabilities of the neural network controllers.

The results of our study can be applied to study the collective motion of self-driving cars, as well as to any collection of small agents that are capable of perceiving their surrounding environment and passing it to a controller. Such ``smart'' agents that are capable of following Markov Decision Process have been called ``smarticles'' or ``smart'' matter. The research field of smarticles is still young, mostly driven by robotics and active matter studies, and yet it has already conceived a few exciting research directions \cite{Cichos2020}. The one that we find particularly interesting focuses on a collection of interacting smart agents that can spontaneously develop unique interaction strategies that emerge in the course of optimization of some unclear cost function. What connections does the resulting emergent behavior share with the winning strategies described by the game theory? This and other interesting questions can be investigated with a ``smart'' version of the predator-prey model, where both prey and predators can learn new ways of hunting or avoiding strategies through deep reinforcement learning or evolutionary algorithms \cite{pezzza_youtube}.

\begin{acknowledgments}
This work would not have been possible without the support of Steffen Rulands and contributions from Jonas Märtens and Jonas Braun, who helped me to shape the vague ideas I had by running computer simulations as a part of their Bachelor Thesis work. Hopefully, as they are now joining the research group that I am leaving, they will take over this work from the point where I am leaving it and submit the follow-up version to a good journal. I also thank Alexander Ziepke, Matteo Ciarchi, and my colleagues from LMU and MPI-PKS for the useful discussions I had with them. 
\end{acknowledgments}

\bibliographystyle{apsrev4-1}
\bibliography{references}

\end{document}